\documentclass[pra,twocolumn,superscriptaddress,aps,floatfix]{revtex4-2}

\PassOptionsToPackage{hyphens}{url}
\usepackage[T1]{fontenc}
\usepackage[utf8]{inputenc}
\usepackage{amsmath}
\usepackage{amssymb}
\usepackage{bm}
\usepackage{graphicx}
\usepackage{braket}
\usepackage{physics}
\usepackage{algpseudocode}
\usepackage{xcolor}
\usepackage{hyperref}
\hypersetup{hidelinks}
\usepackage{orcidlink}

\newcommand{\eref}[1]{Eq.\,\eqref{#1}}
\newcommand{\fref}[1]{Fig.\,\ref{#1}}
\newcommand{\sref}[1]{Sec.\,\ref{#1}}

\newcommand{\bigO}{\mathcal{O}}

\newcommand{\mc}[1]{\mathcal{#1}}

\usepackage{xcolor}
\definecolor{dgreen}{rgb}{0.0,0.6,0.0}
\definecolor{pink}{rgb}{1,0,0.9}

\begin{document}

%\title{Ensemble-rank-truncated propagation for open quantum systems with tridiagonalizable Hamiltonians: {\color{red} near-linear scaling with system size} }

\title{Low-rank propagation for tridiagonalizable open quantum systems: \\ near-linear scaling with system size}

\author{Roman Ovsiannikov\orcidlink{0009-0003-9558-4132}}
\email{roman.ovsiannikov@kipt.kharkov.ua}
\affiliation{Akhiezer Institute for Theoretical Physics, NSC KIPT, Akademichna 1, 61108 Kharkiv, Ukraine}

\author{Kurt Jacobs\orcidlink{0000-0003-0828-6421}}
\email{dr.kurt.jacobs@gmail.com}
\affiliation{Advanced Photonic and Electronic Sciences Division, U.S. Army DEVCOM Army Research Laboratory, Adelphi, Maryland 20783, USA}
\affiliation{Department of Physics, University of Massachusetts at Boston, Boston, Massachusetts 02125, USA}

\author{Andrii G. Sotnikov\orcidlink{0000-0002-3632-4790}}
\email{a\_sotnikov@kipt.kharkov.ua}
\affiliation{Akhiezer Institute for Theoretical Physics, NSC KIPT, Akademichna 1, 61108 Kharkiv, Ukraine}
\affiliation{Education and Research Institute ``School of Physics and Technology'', Karazin Kharkiv National University, Svobody Square 4, 61022 Kharkiv, Ukraine}

\author{Denys I. Bondar\orcidlink{0000-0002-3626-4804}}
\email{dbondar@tulane.edu}
\affiliation{Department of Physics and Engineering Physics, Tulane University, New Orleans, Louisiana 70118, USA}

\date{\today}

\begin{abstract}
The quadratic growth of the density matrix with Hilbert-space dimension $D$ is the central obstacle to simulating large open quantum systems. We introduce a deterministic algorithm that eliminates it for Lindblad dynamics whose Hamiltonian consists of a time-dependent diagonal part plus terms that are tridiagonal after reordering the basis. The state is a low-rank ensemble of vectors, propagated by tridiagonal split-operator steps and ensemble rank truncation of short-time Kraus branches, so that memory and cost per step are linear in $D$ at fixed rank. For a driven nitrogen-vacancy--cavity model, rank 16 reproduces full density-matrix observables to relative error below $10^{-5}$, the method is up to two orders of magnitude faster than QuTiP already at $D\approx 500$, and its runtime scales nearly linearly.
\end{abstract}

\maketitle

\section{Introduction}

Numerical propagation of open quantum systems is frequently required in quantum optics, condensed-matter physics, quantum control, as well as modeling of noisy quantum hardware.  
Even when a Markovian description is adequate, the density operator of a Hilbert space of dimension $D$ contains $D^2$ complex entries, and vectorization leads to an evolution problem in a space of dimension~$D^2$.  
Sparse representations and developed software environments substantially extend the accessible range~\cite{Lambert2026,MercurioEtAl2025}, but they do not remove this quadratic state-storage bottleneck.  
A second challenge is geometric: a useful time integrator should control its error while respecting Hermiticity, positivity, and unit trace.  
These requirements have motivated recent structure-preserving, exponential, and Kraus-form discretizations of the Lindblad equation~\cite{CaoLu2025, ChenEtAl2026, AppeloCheng2025, HuAppeloCheng2025}.

Low-rank representations provide a direct route around density-matrix storage when only a small part of its spectrum is significantly occupied, while observable-adapted model reduction can compress the generator itself~\cite{LeBrisRouchon2013, finazzi_corner-space_2015, 
 chen_low_2020, donatella_continuous-time_2021,GrigolettoEtAl2025}.  
Ensemble rank truncation (ERT) constructs a short-time ensemble from an approximate Kraus map and compresses that ensemble by principal components~\cite{McCaul2021}.  
Recent work has introduced adaptive variational low-rank evolution \cite{GravinaSavona2024}, applied low-rank propagation to large quantum-control models \cite{GoutteSavona2026}, formulated a low-rank variational quantum algorithm \cite{SantosSongSavona2025}, and combined positive low-rank factors with tensor-train compression
\cite{DelMastroAppeloCheng2026}.  
These approaches share a decisive condition:
the useful rank must remain much smaller than $D$ during the time interval of
interest.

Other simulation strategies compress different aspects of the problem.
Stochastic unravelings replace the density matrix by independently propagated
wave functions; adaptive unravelings can reduce trajectory entanglement or sampling variance~\cite{VovkPichler2022, CaoHeLi2026}, and their combination with matrix-product states has enabled large-scale Markovian simulations~\cite{SanderEtAl2025}.  
Tensor-network and process-tensor methods instead exploit restricted operator entanglement, environmental memory, or spatial locality~\cite{LinkTuStrunz2024, HryniukSzymanska2024, JorgensenPollock2019, CygorekEtAl2022, CygorekEtAl2024, FuxEtAl2023, FowlerWrightEtAl2022,
GribbenEtAl2022}.
They can reach system sizes inaccessible to unstructured density-matrix propagation, but their efficiency is controlled by a bond dimension or memory dimension rather than by the density-matrix rank.

Neural and variational representations form a further complementary class.
Restricted-Boltzmann-machine, autoregressive, and deep-network ansatzes have been used for dissipative dynamics and nonequilibrium steady states~\cite{HartmannCarleo2019, NagySavona2019, VicentiniEtAl2019, YoshiokaHamazaki2019, LuoEtAl2022, RehSchmittGaerttner2021, MellakEtAl2024},
while adaptive variational algorithms have also been formulated for quantum processors~\cite{ChenGomesNiuDeJong2024}.  
Their favorable regime is governed by the expressivity and trainability of the chosen ansatz. 
By contrast, the method developed here is deterministic, uses only linear-algebra operations, and targets finite-dimensional models whose generators have an exploitable banded structure.

A separate opportunity comes from the operators rather than the state.
Product formulas reduce an exponential of a sum to exponentials of simpler terms; modern analyses relate their error to nested commutators and locality~\cite{Blanes2024, ChildsSu2019, ChildsEtAl2021}.  
Finite-step Kraus constructions likewise make the order and complete-positivity properties of an open-system update explicit \cite{WonglakhonWisemanChantasri2024}.  
In many light--matter models, diagonal terms coexist with couplings that become tridiagonal after merely reordering the product basis~\cite{Gunderman2025, bondar_symplectic_2026, OvsiannikovSplit2026}.  
Their action can then be evaluated with diagonal multiplications, permutations, and banded linear
solves, without forming a dense Hamiltonian exponential.

In this work, we combine these two reductions.  The coherent evolution is advanced by a symmetric split-operator step for a diagonal plus tridiagonalizable Hamiltonian, while dissipation is represented by ERT branch operators and compressed to a prescribed rank through a small Gram matrix.
The resulting storage is linear in $D$ for a fixed rank and a fixed number of collapse channels.  
We state these qualifications explicitly: the full open-system method is not uniformly efficient for arbitrary Lindblad dynamics, and its accuracy and advantage vanish if the physical state develops a broad spectrum or if the banded operator structure is absent.

The rest of the paper is organized as follows.  In \sref{sec:general_problem} we define the class of open quantum systems considered here.  In \sref{sec:general_algorithm} we describe the tridiagonal split-operator step, the ERT representation of the dissipative map, the rank-truncation procedure, and the scaling estimates.  
In \sref{sec:nv_test} we specialize the construction to a model of a finite  ensemble of nitrogen-vacancy (NV) centers coupled to the cavity modes.
Section~\ref{sec:results} presents comparisons with QuTiP, convergence and timing benchmarks, and the limits of applicability; \sref{sec:conclusion} summarizes the conclusions.

\section{General open-system problem}\label{sec:general_problem}

We consider a finite-dimensional open quantum system governed by the Gorini--Kossakowski--Sudarshan--Lindblad (GKSL) equation
\cite{Gorini1976,Lindblad1976,Jacobs14,WM10}
\begin{equation}\label{eq:general_lindblad}
    \dot{\rho}=-i[H(t),\rho]
    +\sum_{\mu=1}^{K}\left(A_\mu\rho A_\mu^\dagger
    -\frac{1}{2}\{A_\mu^\dagger A_\mu,\rho\}\right),
\end{equation}
where the rates are included in the collapse operators $A_\mu$ and $\hbar=1$.  The Hilbert-space dimension is denoted by $D$, and the number of active collapse channels is $K$.

The algorithm is applicable to Hamiltonians that can be decomposed as
\begin{equation}\label{eq:general_H_decomp}
    H(t)=D_0(t)+\sum_{s=1}^{S}H_s,
\end{equation}
where $D_0(t)$ is diagonal in a reference basis, while each $H_s$ is tridiagonal after a known permutation of that basis.  More explicitly, for every $s$ there is a permutation matrix~$P_s$ such that
\begin{equation}\label{eq:tridiagonalizable}
    T_s=P_s H_s P_s^T
\end{equation}
 is tridiagonal.  The permutations are not dense basis transformations; they only reindex the same basis states.  Therefore, applying $P_s$ or $P_s^T$ to a state vector costs $\bigO(D)$ and can be precomputed once.

For the dissipative part, we assume that each collapse channel has a natural ordering in which the required short-time factors are not computationally costly.  In the implementation used below, $A_\mu$ is tridiagonal in this ordering, $A_\mu^2$ is at most five-banded, and $A_\mu^\dagger A_\mu$ is diagonal or banded.  This assumption can be relaxed to block-tridiagonal or more general banded forms, but the cleanest scaling is obtained for the tridiagonal and five-banded cases.

The density matrix is not stored explicitly. Instead, we use the low-rank ansatz \cite{LeBrisRouchon2013,LeBrisRouchonRoussel2015,McCaul2021}:
\begin{equation}\label{eq:rho_ert}
    \rho \simeq \Psi\Psi^\dagger
    =\sum_{\alpha=1}^{R}\ket{\psi_\alpha}\bra{\psi_\alpha},
\end{equation}
where $\Psi\in\mathbb{C}^{D\times R}$ is a matrix whose columns are unnormalized ensemble vectors.  The integer $R$ is the retained ensemble rank and is treated as a convergence parameter.  In this representation, expectation values are computed as
\begin{equation}\label{eq:expectation_ert}
    \langle O\rangle = \mathrm{Tr}(O\rho)
    \simeq \mathrm{Tr}(\Psi^\dagger O\Psi).
\end{equation}

\section{General tridiagonal ERT algorithm}\label{sec:general_algorithm}

\subsection{Split propagation of the coherent part}

We first describe the coherent Hamiltonian propagator employed as a substep of the full open-system update.  This is the part of the algorithm where the second-order Suzuki--Trotter (or Strang) structure enters explicitly.  Denote the midpoint diagonal operator by
\begin{equation}\label{eq:midpoint_diagonal}
    D_m \equiv D_0(t+\delta t/2).
\end{equation}
For the Hamiltonian decomposition as in \eref{eq:general_H_decomp}, we define a nested symmetric product recursively.  The zeroth product contains only the diagonal part,
\begin{equation}\label{eq:strang_recursive_0}
    \mc{S}_0(\tau)=\exp(-i\tau D_m),
\end{equation}
and the addition of the $r$th tridiagonalizable Hamiltonian term is performed as
\begin{align}\label{eq:strang_recursive_r}
    \mc{S}_r(\tau)
    &=\mc{S}_{r-1}(\tau/2)
      \exp(-i\tau H_r)
      \mc{S}_{r-1}(\tau/2),\notag\\
    &\hspace{4em} r=1,\ldots,S.
\end{align}
The coherent propagator applied over one time step is then
\begin{equation}\label{eq:general_strang}
    U_H(t+\delta t,t)=\mc{S}_S(\delta t)+\bigO(\delta t^3).
\end{equation}
This formula is the explicit palindromic Suzuki--Trotter decomposition \cite{Trotter1959,Strang1968,Suzuki1976,Blanes2024,ChildsEtAl2021}.  The local splitting error of the coherent Hamiltonian part is therefore $\bigO(\delta t^3)$, and the corresponding global error is $\bigO(\delta t^2)$ for a fixed final time.

For the three-term structure used in the test problem below, $H(t)=D_0(t)+H_0+V$, Eqs.~\eqref{eq:strang_recursive_0}--\eqref{eq:general_strang} yield
\begin{equation}\label{eq:nv_strang_generic_section}
\begin{aligned}
U_H(t+\delta t,t)
&=e^{-i\frac{\delta t}{4}D_m}
  e^{-i\frac{\delta t}{2}H_0}
  e^{-i\frac{\delta t}{4}D_m}
  e^{-i\delta t V} \\
&\quad\times
  e^{-i\frac{\delta t}{4}D_m}
  e^{-i\frac{\delta t}{2}H_0}
  e^{-i\frac{\delta t}{4}D_m}
  +\bigO(\delta t^3).
\end{aligned}
\end{equation}
This is the ordering implemented for the Hamiltonian part.  The factors containing $D_m$ are applied as exact element-wise phase multiplications.  The factors containing $H_0$ and $V$ are applied in the basis orderings in which these matrices are tridiagonal; changing between the orderings is only a precomputed permutation of the state-vector entries.

For any tridiagonal or banded matrix \(M\), the elementary exponential needed below is evaluated using the Cayley transform or, equivalently, the \(R_{1,1}\) or \([1/1]\) Padé approximant \cite{Higham2005, AlMohyHigham2011},
\begin{equation}
    e^{cM}x
    \approx
    \left(I-\frac{c}{2}M\right)^{-1}
    \left(I+\frac{c}{2}M\right)x
    +O(c^3).
    \label{eq:cayley_kernel}
\end{equation}
For the coherent Hamiltonian factors, one sets \(c=-i\alpha\) and \(M=T\), where \(T\) is Hermitian and tridiagonal.  In this case, the approximation is unitary up to the accuracy of the linear solve.
For Hermitian $T$, this rational approximation is unitary up to the accuracy of the linear solve.  Since the left-hand side is tridiagonal, \eref{eq:cayley_kernel} is evaluated by the Thomas algorithm or by a banded solver in $\bigO(D)$ operations for a single vector.  For an ERT ensemble with $R$ columns, the same banded matrix is solved with $R$ right-hand sides, giving cost $\bigO(DR)$.  The Cayley approximation has the same local order as the Strang split, so replacing the tridiagonal exponentials in \eref{eq:general_strang} with Cayley transforms does not reduce the coherent step below second order.

If a Hamiltonian term is tridiagonal in a different ordering, the state is first permuted to that ordering, the tridiagonal Cayley transform is applied, and the state is permuted back.  No dense matrix multiplication is involved.  This is the central structural ingredient inherited from the closed-system tridiagonal split-operator method~\cite{OvsiannikovSplit2026}.

\subsection{From the Lindblad equation to ERT branch operators}

We now introduce the branch operators used in the dissipative ERT step. We use \(F_j\) for a generic Kraus operator, \(F^{\rm ERT}_{\mu,\sigma}\) for the full short-time ERT branch, and \(\mathcal{B}_{\mu,\sigma}\) for the reduced dissipative branch that is inserted between two coherent Hamiltonian half-steps.

A general completely positive trace-preserving map can be written in the Kraus form \cite{Kraus1971} as:
\begin{equation}
    \rho(t+\delta t)
    =
    \sum_j F_j \rho(t) F_j^\dagger,
    \quad
    \sum_j F_j^\dagger F_j = 1 .
    \label{eq:kraus_general}
\end{equation}
In the ERT representation the single Kraus label is resolved as
\(j=(\mu,\sigma)\), where \(\mu=1,\ldots,K\) labels the active collapse
operators in Eq.~\eqref{eq:general_lindblad}, and
\(\sigma=\pm1\) labels the two sign branches associated with each channel.
Thus the generic Kraus operator \(F_j\) in Eq.~\eqref{eq:kraus_general}
is replaced by the ERT branch \(F^{\rm ERT}_{\mu,\sigma}\).  Here \(K\)
denotes only the number of active Lindblad channels.

For the Lindblad generator in Eq.~\eqref{eq:general_lindblad}, the full short-time dissipative  ERT branch introduced in Ref.~\cite{McCaul2021} is:
\begin{equation}
    F^{\rm ERT}_{\mu,\sigma}
    =
    \frac{1}{\sqrt{2K}}
    \exp\left[
        \delta t\,G^{\rm full}_\mu
        +
        i\sigma\sqrt{K\delta t}\,A_\mu
    \right],
    \quad
    \sigma=\pm1,
    \label{eq:ert_full_branch}
\end{equation}
with
\begin{equation}
    G^{\rm full}_\mu
    =
    -iH_m
    +
    \frac{K}{2}
    \left(
        A_\mu^2
        -
        A_\mu^\dagger A_\mu
    \right),
    \quad
    H_m \equiv H(t+\delta t/2).
    \label{eq:ert_G_full}
\end{equation}
The prefactor \((2K)^{-1/2}\) averages over the \(2K\) dissipators.  The
terms linear in \(\sqrt{\delta t}\) cancel out after summing over \(\sigma=\pm1\), while the quadratic contribution produces the recycling
term \(A_\mu\rho A_\mu^\dagger\).  
The drift term proportional to \(A_\mu^2-A_\mu^\dagger A_\mu\) removes the unwanted \(A_\mu^2\) contribution from the branch expansion and leaves the Lindblad
anticommutator drift.  Hence,
\begin{equation}
    \rho(t+\delta t)
    =
    \sum_{\mu=1}^{K}
    \sum_{\sigma=\pm1}
    F^{\rm ERT}_{\mu,\sigma}
    \rho(t)
    F^{{\rm ERT}\dagger}_{\mu,\sigma}
    +
    \bigO(\delta t^2),
    \label{eq:ert_kraus_map}
\end{equation}
which reproduces Eq.~\eqref{eq:general_lindblad} to first order in \(\delta t\).

In the split implementation the coherent Hamiltonian evolution is handled separately by the tridiagonal propagator.  We therefore remove the Hamiltonian part from \(G^{\rm full}_\mu\) and keep only the dissipative drift,
\begin{equation}
    G^{\rm diss}_\mu
    =
    \frac{K}{2}
    \left(
        A_\mu^2
        -
        A_\mu^\dagger A_\mu
    \right).
    \label{eq:ert_G_diss}
\end{equation}
The dissipative ERT branch employed in the algorithm is then
\begin{equation}
    \mathcal{B}_{\mu,\sigma}
    =
    \frac{1}{\sqrt{2K}}
    \exp\left[
        \delta t\,G^{\rm diss}_\mu
        +
        i\sigma\sqrt{K\delta t}\,A_\mu
    \right].
    \label{eq:ert_diss_branch_compact}
\end{equation}
This operator is not the full Kraus operator of the time step.  It is only
the dissipative branch inserted between the two coherent half-steps.  In the numerical implementation, the scalar factor \((2K)^{-1/2}\) can be applied separately as a
branch weight.

The corresponding Kraus branch of the split one-step map is
\begin{equation}
    F^{\rm split}_{\mu,\sigma}(t+\delta t,t)
    =
    U_H(t+\delta t,t+\delta t/2)\,
    \mathcal{B}_{\mu,\sigma}(\delta t)\,
    U_H(t+\delta t/2,t).
    \label{eq:split_kraus_operator}
\end{equation}
Before rank truncation, the split update is therefore
\begin{equation}
    \rho(t+\delta t)
    =
    \sum_{\mu=1}^{K}
    \sum_{\sigma=\pm1}
    F^{\rm split}_{\mu,\sigma}
    \rho(t)
    F^{{\rm split}\dagger}_{\mu,\sigma}
    +
    \bigO(\delta t^2).
    \label{eq:split_density_kraus}
\end{equation}
The ERT-specific approximation then enters through the compression of the
enlarged set of branches back to the retained rank \(R\).

\subsection{Banded realization of the branch operators}\label{sec:banded_branch}

Equation~\eqref{eq:ert_diss_branch_compact} gives the compact formal
origin of the dissipative branch \(\mathcal{B}_{\mu,\sigma}\).  In the
default numerical implementation this compact exponential is not formed
as a dense matrix exponential.  Instead, it is evaluated as the
factorized banded product
\begin{align}\label{eq:ert_diss_branch_factored}
    \mathcal{B}_{\mu,\sigma}
    &\approx \frac{1}{\sqrt{2K}}
    \exp\left(i\sigma\sqrt{K\delta t}\,A_\mu\right)\notag\\
    &\quad\times
    \exp\left(\frac{K\delta t}{2}A_\mu^2\right)
    \exp\left(-\frac{K\delta t}{2}A_\mu^\dagger A_\mu\right) .
\end{align}

The factorized form in Eq.~\eqref{eq:ert_diss_branch_factored} should
be understood as a weak-dissipation and small-time-step realization of
the compact branch in Eq.~\eqref{eq:ert_diss_branch_compact}.  The
small parameters controlling this replacement are the dimensionless
one-step dissipative factors, e.g.,
\[
    K\delta t\,\|A_\mu^\dagger A_\mu\| \ll 1,
    \qquad
    K\delta t\,\|A_\mu^2\| \ll 1 .
\]
In this regime, the damping and \(A_\mu^2\) factors are close to the identity, and the additional commutator terms generated by replacing the compact exponential with the ordered product are local higher-order corrections to the infinitesimal ERT map.  
This is the regime targeted by the present implementation and by the numerical tests below.  For stronger dissipative rates, one must reduce the time step, increase the retained rank and check convergence, or evaluate the branch factors in a less approximate mode.

The difference between the compact and factored forms is therefore a local splitting error coming from the noncommutativity of the factors.
The ordered product is useful because the three factors have simple sparsity patterns in the applications considered below. 
If
\(A_\mu\) is tridiagonal in its natural ordering, then \(A_\mu^2\) is five-banded, and \(A_\mu^\dagger A_\mu\) is often diagonal or at most
banded.  Unlike the coherent propagator in
Eq.~\eqref{eq:general_strang}, this dissipative branch factorization is
not a second-order Suzuki--Trotter decomposition of the Lindblad
dissipator.  It is the practical banded realization of the original
infinitesimal ERT Kraus branch.  
We therefore regard this dissipative step as a first order short-time ERT update: before rank truncation, it is designed to reproduce the Lindblad generator to first order in \(\delta t\), with a local error of order \(O(\delta t^2)\) in the same sense as the original infinitesimal ERT construction, plus the additional local splitting error introduced by the ordered banded factorization.

The diagonal damping factor is computationally cheap because \(A_\mu^\dagger A_\mu\) is diagonal in the applications considered here.  
It can therefore be applied either exactly as an element-wise exponential or, in the weak-damping regime, by a linear or quadratic polynomial approximation.
The \(A_\mu^2\) factor is five-banded and is applied either by a banded Cayley solve using Eq.~\eqref{eq:cayley_kernel}, or by the same weak-step polynomial approximation. 
Finally, the factor containing \(A_\mu\) is applied by Eq.~\eqref{eq:cayley_kernel} in the ordering where \(A_\mu\) is tridiagonal.

This is the open-system analogue of the tridiagonal Cayley step used
for the Hamiltonian factors.  The important point is that the
application of every branch operator to an ensemble vector requires only banded matrix-vector products and banded linear solves.

Given an ensemble matrix $\Psi\in\mathbb{C}^{D\times R}$, the uncompressed branch ensemble is
\begin{equation}\label{eq:general_branch_ensemble}
    \Phi=\left[\mathcal{B}_{1,-}\Psi,\mathcal{B}_{1,+}\Psi,
    \ldots,\mathcal{B}_{K,-}\Psi,\mathcal{B}_{K,+}\Psi\right],
\end{equation}
so that $\Phi\in\mathbb{C}^{D\times 2KR}$.  If this matrix were kept without truncation, the rank would grow by a factor $2K$ at every step.  The compression stage described below is what turns the Kraus branching into a practical low-rank algorithm.

\subsection{Compression by the Gram matrix}

After the dissipative branching step, the enlarged ensemble matrix
\(\Phi\) represents the density matrix
\begin{equation}
    \rho_\Phi
    =
    \Phi\Phi^\dagger .
    \label{eq:branch_density}
\end{equation}
The number of columns of \(\Phi\) is \(2KR\), because each of the
\(R\) incoming ensemble vectors is mapped into two branches for each
of the \(K\) collapse channels.  Keeping all these columns would make
the rank grow by a factor \(2K\) at every time step.  The purpose of
the ERT compression step is to replace \(\Phi\) by a new ensemble
\(\Psi_{\rm new}\in\mathbb{C}^{D\times R}\) that gives the best
rank-\(R\) approximation within the span of the generated branch
vectors.

We do this without forming the \(D\times D\) density matrix.  First,
we form the small overlap matrix, also known as the Gram matrix,
\begin{equation}
    S
    =
    \Phi^\dagger \Phi .
    \label{eq:gram_matrix}
\end{equation}
This matrix has dimension \(2KR\times 2KR\), independent of the
Hilbert-space dimension \(D\) except through the dot products used to
form it.  The entries of \(S\) are the inner products between all
branch vectors in \(\Phi\), thus \(S\) contains all information needed to orthogonalize the ensemble inside its current span.

Let the eigendecomposition of the Gram matrix be
\begin{equation}
    V^\dagger S V
    =
    \operatorname{diag}(w_1,w_2,\ldots),
    \qquad
    w_1\geq w_2\geq \cdots ,
    \label{eq:gram_eigh}
\end{equation}
where the columns \(v_j\) of \(V\) are the eigenvectors of \(S\).
We define the rotated ensemble as
\begin{equation}
    \bar{\Psi}
    =
    \Phi V .
    \label{eq:rotated_branch_ensemble}
\end{equation}
Its overlap matrix is diagonal:
\begin{equation}
    \bar{\Psi}^\dagger \bar{\Psi}
    =
    V^\dagger \Phi^\dagger \Phi V
    =
    V^\dagger S V
    =
    \operatorname{diag}(w_1,w_2,\ldots).
    \label{eq:rotated_overlap}
\end{equation}
Thus the columns of \(\bar{\Psi}\) are mutually orthogonal
unnormalized ensemble vectors, and the squared norm of the \(j\)-th
column is \(w_j\).  In other words, diagonalizing the small Gram
matrix is equivalent to finding the principal components of the branch
ensemble \(\Phi\), but it avoids diagonalizing the full density matrix
\(\rho_\Phi\).

The rotation by \(V\) does not change the represented density matrix,
because \(V\) is unitary:
\begin{align}
    &\rho_\Phi
    =
    \Phi\Phi^\dagger
    =
    \Phi V V^\dagger \Phi^\dagger
    =
    \bar{\Psi}\bar{\Psi}^\dagger
    =
    \sum_j
    \ket{\bar{\psi}_j}\bra{\bar{\psi}_j},
    \notag\\
    &\ket{\bar{\psi}_j}=\Phi v_j .
    \label{eq:branch_density_rotated}
\end{align}
The truncation is therefore performed only after rotating to this
orthogonal principal-component basis.  If \(V_R=(v_1,\ldots,v_R)\)
contains the first \(R\) eigenvectors, corresponding to the \(R\)
largest weights \(w_j\), the compressed ensemble is
\begin{equation}
    \Psi_{\rm new}
    =
    \Phi V_R .
    \label{eq:compressed_ensemble}
\end{equation}
Equivalently, Eq.~\eqref{eq:compressed_ensemble} is the truncated
singular-value decomposition of the branch matrix \(\Phi\), and hence a best rank-$R$ approximation in the Frobenius and spectral norms by the Eckart-Young theorem \cite{EckartYoung1936}. The discarded weight \(\sum_{j>R}w_j\) is a natural local diagnostic of the rank-truncation error. After compression, the ensemble can be trace-normalized by rescaling \(\Psi_{\rm new}\), so that
\begin{equation}
    \operatorname{Tr}
    \left(
        \Psi_{\rm new}\Psi_{\rm new}^\dagger
    \right)
    =
    1 .
    \label{eq:trace_normalization}
\end{equation}

In the wavefunction form, Eq.~\eqref{eq:split_kraus_operator} means that
each current ensemble member is first propagated by the Hamiltonian
half-step, then branched by all dissipative operators
\(\mathcal{B}_{\mu,\sigma}\), and finally all branches are compressed
back to rank \(R\).  The last Hamiltonian half-step can be applied
after the compression because it is common to all branches and
therefore does not change the Gram matrix used for the
principal-component truncation.

The complete split step for the ensemble can therefore be written as
\begin{align}
    &\Psi_{1/2}
    =
    U_H(t+\delta t/2,t)\Psi(t),
    \nonumber\\
    &\Phi_{1/2}
    =
    \left[
    \mathcal{B}_{1,-}\Psi_{1/2},
    \mathcal{B}_{1,+}\Psi_{1/2},
    \ldots,
    \mathcal{B}_{K,-}\Psi_{1/2},
    \mathcal{B}_{K,+}\Psi_{1/2}
    \right],
    \nonumber\\
    &\widetilde{\Psi}_{1/2}
    =
    \mathcal{T}_R[\Phi_{1/2}],
    \nonumber\\
    &\Psi(t+\delta t)
    =
    U_H(t+\delta t,t+\delta t/2)\widetilde{\Psi}_{1/2}.
    \label{eq:full_split_step}
\end{align}
Here \(\mathcal{T}_R\) denotes the Gram-matrix truncation described in
Eq.~\eqref{eq:compressed_ensemble}.  The half-step propagators are
obtained from the same Suzuki--Trotter construction as
Eq.~\eqref{eq:general_strang}, with \(\delta t\) replaced by
\(\delta t/2\) and with the appropriate midpoint time.

\subsection{Scaling}

For fixed $R$ and fixed number of collapse channels $K$, all persistent arrays scale linearly with $D$ \cite{McCaul2021,AppeloCheng2025,ChenEtAl2026}. The ensemble itself requires $\bigO(DR)$ memory, and the temporary branch matrix requires $\bigO(DKR)$ memory.  Each diagonal, tridiagonal, or five-banded operation on the ensemble is linear in $D$ for fixed rank.  The most important dense operation is the ERT compression, which forms and diagonalizes the $2KR\times 2KR$ Gram matrix.  Its cost is
\begin{equation}\label{eq:general_cost}
    \bigO\left[D(2KR)^2\right]+\bigO\left[(2KR)^3\right].
\end{equation}
Thus, the formal dependence on Hilbert-space dimension remains linear when $K$ and $R$ are fixed.  The rank dependence is stronger; therefore, $R$ must be chosen by convergence testing for the particular physical regime.

\section{NV-center spin--cavity system as a test problem}\label{sec:nv_test}

We now apply the general construction to a finite driven spin--cavity model motivated by active research on realizations of NV-center ensembles coupled to microwave resonators \cite{Doherty2013,Kubo2010,Amsuss2011,OvsiannikovHybrid2026,OvsiannikovOptimal2026}.  This section is a test of the algorithm, not a prerequisite for its formulation.

The cavity annihilation operator is $a$, and the collective spin operators are $J_z$, $J_-$, and $J_+=J_-^\dagger$.  The finite-dimensional Hamiltonian is
\begin{equation}\label{eq:nv_H}
    H(t)=\omega_c a^\dagger a + \Delta(t)J_z
    + g(a+a^\dagger)(J_-+J_+),
\end{equation}
with
\begin{equation}\label{eq:nv_Delta}
    \Delta(t)=\omega_{s0}+\Lambda\sin(\omega_{\rm mod}t+\phi).
\end{equation}
The Hilbert-space dimension is
\begin{equation}\label{eq:nv_dimension}
    D=N_{\rm cav}(2J+1),
\end{equation}
where $N_{\rm cav}$ is the photon cutoff and $J$ is the collective spin size.

The Hamiltonian is decomposed as
\begin{equation}\label{eq:nv_decomp}
    H(t)=H_0+V+D_0(t),
\end{equation}
where
\begin{align}
    H_0 &= g(aJ_+ + a^\dagger J_-),\label{eq:nv_H0}\\
    V &= g(aJ_- + a^\dagger J_+),\label{eq:nv_V}\\
    D_0(t) &= \omega_c a^\dagger a + \Delta(t)J_z.\label{eq:nv_D0}
\end{align}
The diagonal term $D_0(t)$ is diagonal in the product basis $\ket{n}\otimes\ket{J,m}$.  The term $H_0$ conserves $n+m$, so ordering the basis by $n+m$ makes it tridiagonal.  The term $V$ conserves $n-m$, so ordering the basis by $n-m$ makes it tridiagonal.  The transformation between these two orderings is a precomputed permutation.

For the dissipative test, we use the Lindblad channels
\begin{align}
    A_1&=\sqrt{\gamma_{\rm cav}(1+n_T)}\,a,
    &A_2&=\sqrt{\gamma_{\rm cav}n_T}\,a^\dagger,\label{eq:nv_cavity_channels}\\
    A_3&=\sqrt{\kappa_{\rm spin}(1+n_s)}\,J_-,
    &A_4&=\sqrt{\kappa_{\rm spin}n_s}\,J_+.
    \label{eq:nv_spin_channels}
\end{align}
The cavity ladder operators are tridiagonal when the basis is ordered by photon number within each spin sector, while the spin ladder operators are tridiagonal when the basis is ordered by $m$ within each photon sector.  Their squares are five-banded, and the products $a^\dagger a$, $aa^\dagger$, $J_+J_-$, and $J_-J_+$ are diagonal.  Therefore this model exactly matches the operator assumptions of \sref{sec:general_algorithm}.

The monitored cavity quadratures are
\begin{equation}\label{eq:nv_quadratures}
    X=\frac{a+a^\dagger}{2},\qquad
    P=\frac{a-a^\dagger}{2i},
\end{equation}
and we compare the variances
\begin{equation}\label{eq:nv_variances}
    \mathrm{Var}(X)=\langle X^2\rangle-\langle X\rangle^2,
    \qquad
    \mathrm{Var}(P)=\langle P^2\rangle-\langle P\rangle^2.
\end{equation}
These observables were chosen because they are sensitive to the same amplification and squeezing physics studied in the linearized NV-cavity model, while still being computable in the finite-spin density-matrix simulation.

\section{Numerical results}\label{sec:results}

\subsection{Direct comparison with QuTiP}

For the direct ERT--QuTiP comparison, we use $N_{\mathrm{cav}}=15$ and $J=5$, giving $D=165$. The numerical parameters, in the units used by the simulations, are $\omega_c=2\pi\times2.4 ~GHz$, $\omega_{s0}=2\pi\times3.6~GHz$, $\omega_{\mathrm{mod}}=2\pi\times6.0~GHz$, $\Lambda=2\pi\times0.5~GHz$, $g=2\pi\times0.01/\sqrt{2J}~GHz$, $\gamma_{\mathrm{cav}}=\kappa_{\mathrm{spin}}=2\times10^{5}~ s^{-1}$, $n_T=0.05$, $n_s=0.02$, and $\phi=0$. The evolution interval is $0\le t\le5$~ns, with an ERT time step $\delta t=1/720$~ns and 3601 uniformly spaced output times.

\begin{figure}[t]
    \centering
    \includegraphics[width=\columnwidth]{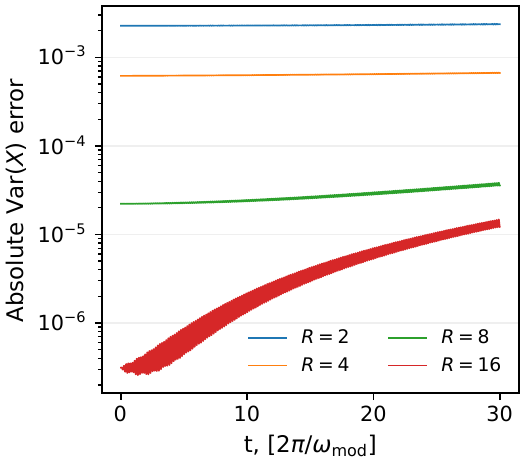}
    \caption{Absolute error in the cavity quadrature variance, $|\operatorname{Var}_{\mathrm{ERT},R}(X;t) -\operatorname{Var}_{\mathrm{QuTiP}}(X;t)|$, for $R=2,4,8,16$ and $N_{\mathrm{cav}}=15$, $J=5$ ($D=165$). The evolution interval is $0\le t\le5 ~ ns$, with $\delta t=1/720$~ns. The vertical axis is logarithmic. QuTiP starts from the full finite-dimensional thermal product state, whereas ERT starts from its normalized rank-$R$ approximation; the error therefore includes initial-state truncation.}
    \label{fig:varx}
\end{figure}

QuTiP starts from the full normalized thermal product state in the finite Hilbert space \cite{Johansson2012,Johansson2013,Lambert2026}. ERT retains its $R$ largest eigenvalues and renormalizes the retained state:
\begin{equation}
\rho_R(0) = \frac{ \sum_{j=1}^{R}\lambda_j|v_j\rangle\langle v_j| }{ \sum_{j=1}^{R}\lambda_j},\qquad \lambda_1\ge\lambda_2\ge\cdots.
\end{equation}
The comparison therefore includes both initial-state truncation and errors accumulated during propagation. The ERT results in \fref{fig:varx}  use $R=2,4,8,16$, with trace normalization after each complete step.

The calculation uses the finite-spin operators and the factorized banded dissipative update with collapse operators
\begin{align}
A_1 &= \sqrt{\gamma_{\mathrm{cav}}(1+n_T)}\,a, \\ A_2 &= \sqrt{\gamma_{\mathrm{cav}}n_T}\,a^\dagger,
\\ A_3 &= \sqrt{\kappa_{\mathrm{spin}}(1+n_s)}\,J_-, \\
A_4 &= \sqrt{\kappa_{\mathrm{spin}}n_s}\,J_+.
\end{align}
Here $A_1$ and $A_2$ describe cavity photon loss and thermal excitation, respectively, while $A_3$ and $A_4$ describe collective spin relaxation and thermal excitation.

For this configuration, the exponential factors $\exp[-K\delta t\,A_\mu^\dagger A_\mu/2]$ and $\exp[K\delta t\,A_\mu^2/2]$ use first-degree Taylor approximations for the cavity channels $A_1,A_2$ and the spin-excitation channel $A_4$. Second-degree Taylor approximations are used for the spin-relaxation channel $A_3$. The factors $\exp[i\sigma\sqrt{K\delta t}\,A_\mu]$ are evaluated using Cayley transforms. The QuTiP 5.2.3 reference uses absolute and relative tolerances $10^{-9}$ and $10^{-8}$, respectively.

Figure~\ref{fig:varx} resolves the absolute error in $\operatorname{Var}(X)$ for the four retained ranks. The maximum sampled errors over $0\le t\le5$~ns are approximately $2.39\times10^{-3}$, $6.70\times10^{-4}$, $3.83\times10^{-5}$, and $1.47\times10^{-5}$ for $R=2,4,8,16$, respectively. For $R=16$, the initial error is approximately $3.13\times10^{-7}$ and increases during the evolution.

\subsection{Rank dependence of the error}

To quantify convergence with the retained rank, we compute an integrated relative error for each monitored observable,
\begin{equation}\label{eq:integrated_error}
    \epsilon_f =
    \left[
    \frac{\int_0^{t_f} |f_{\rm ERT}(t)-f_{\rm ref}(t)|^2dt}
         {\int_0^{t_f} |f_{\rm ref}(t)|^2dt + \epsilon_0}
    \right]^{1/2}.
\end{equation}
The four observable errors are combined as
\begin{align}
\epsilon_{\mathrm{RMS}} = \left[ \frac{1}{4}\sum_{f\in\mathcal{F}}\epsilon_f^2 \right]^{1/2}, \nonumber \\ 
\mathcal{F} = \left\{ \operatorname{Var}(X), \operatorname{Var}(P), \langle a^\dagger a\rangle, \langle J_z\rangle \right\}.
\end{align}
The integrals are evaluated using the trapezoidal rule. The supplied implementation uses $\epsilon_0=10^{-30}$.

\begin{figure}[t]
    \centering
    \includegraphics[width=\columnwidth]{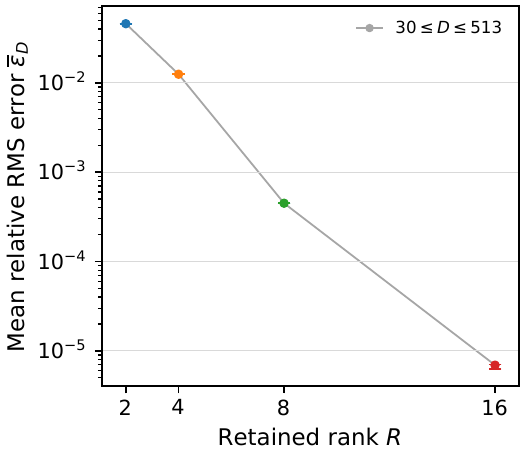}
    \caption{Integrated relative RMS error versus retained rank $R$, combining the errors in $\operatorname{Var}(X)$, $\operatorname{Var}(P)$, $\langle a^\dagger a\rangle$, and $\langle J_z\rangle$ over $0\le t\le 0.5$~ns. Filled circles show results for eight dimensions in the range $30\le D\le513$. Vertical bars indicate the minimum and maximum errors across these dimensions at fixed $R$.}
    \label{fig:error}
\end{figure}

Figure~\ref{fig:error} shows the integrated error as a function of rank over $0\le t\le 0.5~ns$, with 361 output times and $\delta t=1/720$~ns. The mean relative error uses $J=2,3,\ldots,9$ and $N_{\mathrm{cav}}=3J$, corresponding to $D=30,63,108,165,234,315,408,513$. 

The error decreases from approximately $4.6\times10^{-2}$ at $R=2$ to $(6.3\text{--}7.0)\times10^{-6}$ at $R=16$. The relative variation across the primary dimensions, defined as $(\epsilon_{\max}-\epsilon_{\min})/\epsilon_{\min}$, is below $0.6\%$ for $R=2,4,8$ and approximately $11.5\%$ for $R=16$. The dependence on rank is therefore substantially stronger than the dependence on dimension in this benchmark.

The shorter integration interval distinguishes this error measure from the time-resolved comparison in \fref{fig:varx} Moreover, changing $J$ changes the physical spin size. Rank convergence, time-step convergence, and cavity-cutoff convergence at fixed $J$ must therefore be assessed separately.

\subsection{Timing benchmarks}

\begin{figure}[t]
    \centering
    \includegraphics[width=\columnwidth]{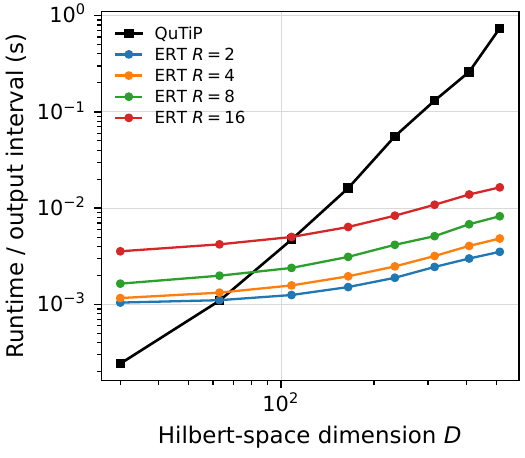}
    \caption{Recorded computation time per output interval for QuTiP and ERT with $R=2,4,8,16$, over $30\le D\le513$. The plotted quantity is the recorded runtime divided by $N_t-1=360$, for $0\le t\le 0.5$~ns and output spacing $\delta t=1/720$~ns. Both axes are logarithmic. }
    \label{fig:time_qutip}
\end{figure}

Figure~\ref{fig:time_qutip} compares the recorded computation time divided by the number of output intervals, 
\begin{equation} 
\tau_{\mathrm{out}} = \frac{T_{\mathrm{recorded}}}{N_t-1}, \qquad N_t-1=360.
\end{equation}
At $D=30$ and $D=63$, QuTiP has the smaller recorded time for all four ranks; at $D\ge165$, all four ERT curves lie below the QuTiP curve. At $D=513$, the times per output interval are approximately $0.739\,\mathrm{s}$ for QuTiP and $0.00352$, $0.00484$, $0.00825$, and
$0.0164\,\mathrm{s}$ for ERT with $R=2,4,8,16$.

\begin{figure}[t]
    \centering
    \includegraphics[width=\columnwidth]{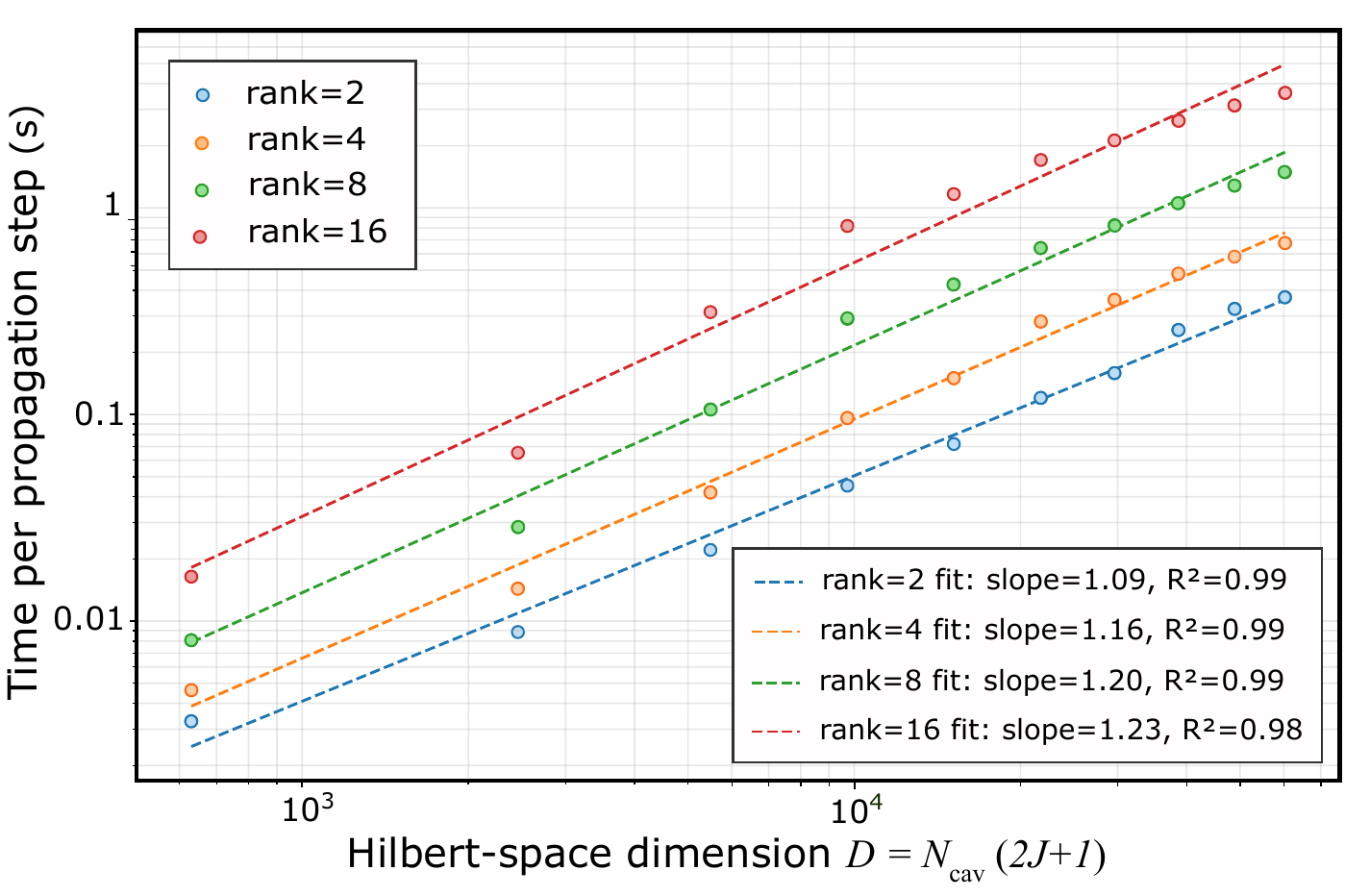}
    \caption{ERT propagation time per physical time step for $R=2,4,8,16$. The reported time is the median duration of three 360-step trajectories divided by 360. Dashed lines show power-law fits using all ten dimensions in the range $630\le D\le60300$. The fitted exponents are $1.092$, $1.157$, $1.198$, and $1.229$ for $R=2,4,8,16$, respectively.}
    \label{fig:only_ert_time}
\end{figure}

Figure~\ref{fig:only_ert_time} presents a separate ERT-only timing series with $J=10,20,\ldots,100$ and $N_{\mathrm{cav}}=3J$, covering $630\le D\le60300$. Each trajectory contains 360 physical propagation steps with $\delta t=1/720$~ns. The reported time is the median of three trajectory durations divided by 360.

Unweighted linear fits of $\ln\tau_R$ against $\ln D$, using all ten dimensions, yield exponents $1.092$, $1.157$, $1.198$, and $1.229$ for $R=2,4,8,16$. The corresponding coefficients of determination in logarithmic space are $0.991$, $0.993$, $0.987$, and $0.978$. These results support near-linear empirical scaling at fixed rank over the tested range.

\subsection{Range of applicability and failure modes}\label{sec:limitations}

The favorable regime of the algorithm is determined jointly by the operator structure and numerical rank.  The linear dependence on $D$ in Eq.~\eqref{eq:general_cost} holds only when $R$ and $K$ remain fixed. If $M=2KR$ denotes the number of branch vectors before compression, a useful low-rank regime requires $R\ll D$ and, in practice, $M\ll D$. Otherwise, the $D\times M$ branch matrix and the $M\times M$ Gram problem remove the memory and runtime advantage.  The same is true if a large number of collapse channels is needed, since the leading compression cost is quadratic in $KR$. Finally, diagonal, tridiagonal, or narrow-banded actions must be available in known orderings; generic dense couplings change the dominant operation from a banded solve to dense linear algebra.  Thus, near-linear scaling at fixed $R$ is not by itself a guarantee of a speed-up at fixed accuracy \cite{McCaul2021,GravinaSavona2024,AppeloCheng2025}.

The Gram spectrum provides an inexpensive diagnostic of whether the rank assumption remains valid.  At step $n$, let $w_j^{(n)}$ be the ordered eigenvalues in Eq.~\eqref{eq:gram_eigh}.  We monitor the relative discarded weight and the effective rank,
\begin{equation}\label{eq:rank_diagnostics}
 \eta_R^{(n)}=
 \frac{\sum_{j>R}w_j^{(n)}}{\sum_j w_j^{(n)}},
 \qquad
 r_{\rm eff}^{(n)}=
 \frac{\left(\sum_jw_j^{(n)}\right)^2}
      {\sum_j\left(w_j^{(n)}\right)^2}.
\end{equation}
A small $\eta_R^{(n)}$ is necessary but not sufficient for global accuracy: discarded components can influence later dynamics, and trace renormalization can conceal their accumulated weight.  Results should therefore be accepted only after the observables of interest are stable under $R\to2R$, while the pre-normalization trace loss, $\max_n\eta_R^{(n)}$, and $r_{\rm eff}^{(n)}$ remain controlled.  These tests must be performed independently of convergence in the Hilbert-space cutoff.

Time-step convergence is a separate requirement.  Practical weak-step indicators for the factorized branch include
\begin{equation}\label{eq:weak_step_indicator}
 \chi_\mu=\max\!\left\{
 \sqrt{K\delta t}\,\|A_\mu\|,
 K\delta t\,\|A_\mu^2\|,
 K\delta t\,\|A_\mu^\dagger A_\mu\|
 \right\}.
\end{equation}
When any $\chi_\mu$ is not small, the ordered branch factorization and its rational or polynomial factors must be checked by reducing $\delta t$.  More generally, the infinitesimal ERT map has local error $\bigO(\delta t^2)$ and hence only first-order global time accuracy before rank truncation.  The second-order coherent Strang--Cayley substep does not raise the order of the full dissipative integrator, and the ordered banded factorization can add commutator error.  A comparison between $\delta t$ and $\delta t/2$ is therefore mandatory; higher-order finite-step Kraus schemes
are a possible extension \cite{WonglakhonWisemanChantasri2024,CaoLu2025,
ChenEtAl2026,HuAppeloCheng2025}.

Equation~\eqref{eq:general_lindblad} also limits the present formulation to time-local Markovian GKSL dynamics.  Non-Markovian models require an enlarged Markovian embedding or a different representation, such as process-tensor methods \cite{StrathearnEtAl2018,JorgensenPollock2019,CygorekEtAl2022}.
Driven lasing systems are a representative unfavorable low-rank regime: continuous pumping, loss, gain saturation, and phase diffusion can produce a slowly decaying density-matrix spectrum, so that the rank required for accurate observables grows during the evolution. This behavior occurred in our separate lasing-system stress test: increasing $R$ reduced the observable errors, but the rank required for acceptable accuracy removed the runtime advantage over the full-density QuTiP reference.  We therefore do not regard ERT as a universal lasing solver.  It is suitable only when the rank and time-step diagnostics above remain favorable for the specific lasing regime.

\section{Conclusion}\label{sec:conclusion}

We have formulated a general ERT propagation algorithm for open quantum systems whose unitary and dissipative generators are diagonal, tridiagonal, or low-bandwidth in suitable basis orderings.  The unitary part is propagated by an explicit second-order Suzuki--Trotter split using diagonal phase multiplications, permutations, and Cayley transforms for tridiagonal matrices.  The dissipative part is represented by the original infinitesimal ERT branch map rather than by a second-order Suzuki--Trotter decomposition of the dissipator: each collapse channel generates two short-time branches and the enlarged ensemble is compressed back to a fixed rank using the Gram matrix.

The finite driven NV-center spin--cavity model was used as a test case because it naturally has the required structure.  The Hamiltonian separates into a diagonal part and two terms that are tridiagonal in different orderings, while the cavity and spin-ladder collapse operators are tridiagonal and their squares are five-banded.  In the tested parameter regime, the ERT propagator reproduces the QuTiP quadrature-variance evolution, shows systematic convergence with rank, and displays near-linear empirical scaling of the pure propagation step with Hilbert-space dimension.

The main limitation is that the retained rank is problem dependent.  Strongly mixed states, long evolution times, or regimes with many strongly populated dissipative branches may require a larger rank.  The second limitation is structural: if additional Hamiltonian or collapse terms destroy either diagonal, tridiagonal, or banded form, the speed-up is reduced unless one finds new orderings or extends the method to block-banded operators.  Nevertheless, for systems with the required sparse structure, the method provides a practical route to finite-dimensional dissipative simulations beyond both closed-system unitary propagation and linearized covariance dynamics.

\section*{Code availability}

The implementation used for the numerical tests is available in repository~\cite{ERTPropCode}.

\acknowledgments
D.I.B. was sponsored by the Army Research Office and was accomplished under Cooperative Agreement Number W911NF-26-2-A181. The views and conclusions contained in this document are those of the authors and should not be interpreted as representing the official policies, either expressed or implied, of the Army Research Office or the U.S. Government. The U.S. Government is authorized to reproduce and distribute reprints for Government purposes notwithstanding any copyright notation herein. R.O. acknowledges support provided by Ukraine to research laboratories/groups of young scientists of the NAS of Ukraine for research in priority areas of science and technology (Grant No. 26/01-2026(7), RC 0126U002200). 

\bibliography{ert_algorithm_refs}

\end{document}